\documentclass[aps,pre,reprint,groupedaddress]{revtex4-1}

\pdfoutput=1
\usepackage{graphicx}
\usepackage{amsmath,bm,dsfont}
\usepackage{siunitx}
\usepackage{amssymb}
\usepackage{mathrsfs}
\usepackage{color}

\newcommand{\so}
   {\mathrel{\rlap{\raise1.3pt\hbox{$>$}}{\lower3pt\hbox{$\sim$}}}}
\newcommand{\io}
   {\mathrel{\rlap{\raise1.3pt\hbox{$<$}}{\lower3pt\hbox{$\sim$}}}}

\begin{document}


\title{Saffman-Delbr\"uck and beyond: a pointlike approach}

\author{Quentin Goutaland and Jean-Baptiste Fournier}

\affiliation{Laboratoire ``Mati\`ere et Syst\`emes Complexes" (MSC), UMR 7057 CNRS, Universit\'e de Paris, 75205 Paris Cedex 13, France}

\date{\today}

\begin{abstract}
We show that a very good analytical approximation of Saffman-Delbr\"uck's (SD) law (mobility of a bio-membrane inclusion) can be obtained  easily from the velocity field produced by a pointlike force in a 2D fluid embedded in a solvent, by using a small wavelength cutoff of the order of the particle's radius~$a$. With this method, we obtain analytical generalizations of the SD law that take into account the bilayer nature of the membrane and the intermonolayer friction $b$. We also derive, in a calculation that consistently couples the quasi-planar two-dimensional (2D) membrane flow with the 3D solvent flow, the correction to the SD law arising when the inclusion creates a local spontaneous curvature. For an inclusion spanning a flat bilayer, the SD law is found to hold simply upon replacing the 2D viscosity $\eta_2$ of the membrane by the sum of the monolayer viscosities, without influence of $b$ as long as $b$ is above a threshold in practice well below known experimental values.
For an inclusion located in only one of the two monolayers (or adhering to one monolayer), the SD law is influenced by $b$ when $b<\eta_2/(4a^2)$. In this case, the mobility can be increased by up to a factor of two, as the opposite monolayer is not fully dragged by the inclusion. For an inclusion creating a local spontaneous curvature, we show that the total friction is the sum of the SD friction and that due to the pull-back created by the membrane deformation, a point that was assumed without demonstration in the literature.
\end{abstract}

\maketitle 

\section{Introduction}
\label{intro}

In the presence of friction, mobility links the velocity of a body to the force applied to it. In three dimensions (3D), at low Reynolds numbers, Stokes law states that the mobility of a spherical particle of radius $a$, in a fluid of viscosity $\eta$, is given by $\mu_{3D}=1/(6\pi\eta a)$~\cite{Lamb_book}.
In 2D, however, the mobility diverges, which is known as the Stokes paradox~\cite{Lamb_book}.
This has to do with the fact that the Oseen tensor~\cite{Lamb_book}, which gives the velocity field conjugated to a point-like force, has an infrared divergence when integrated over wavevectors.
This divergence can be regularized in two different ways: either by restricting the 2D fluid to a finite area, or by embedding the infinite 2D fluid in an immisible 3D fluid~\cite{SaffmanPNAS75,SaffmanJFM76}. This happens naturally in biological membranes, which are 2D fluids of lipids embedded in bulk water. Saffman-Delbr\"uck's (SD) law then tells us that the mobility of a disc of radius $a$ within the membrane is given by $\mu=(4\pi\eta_2)^{-1}\{\ln\,[\eta_2/(\eta a)]-\gamma\}$, where $\eta_2$ is the 2D viscosity of the membrane, $\eta=\frac12(\eta^++\eta^-)$ is half the sum of the viscosities of the solvent phases above and below the membrane, and $\gamma\simeq0.577$ is Euler's constant~\cite{SaffmanPNAS75,SaffmanJFM76}. The finite Saffman-Delbr\"uck length $\ell=\eta_2/(2\eta)$ regularizes the infrared divergence mentioned above. 

As far as we know, there is no easy way to demonstrate the SD law. Available demonstrations require heavy calculations~\cite{SaffmanPNAS75,SaffmanJFM76,HughesJFM81,StoneJFM98,StoneJFM15}. In this paper, we present a very simple derivation, based on a pointlike calculation regularized
by a sharp high wavevector cutoff of the order of the inverse of the particle size. This derivation is not exact, because the dimension of the inclusion is only taken into account up to a multiplicative factor of order unity. However, since the particle's radius appears within a logarithm in the SD law, it turns out to be excellent.
Pointlike approximations are standard in soft matter to calculate interactions among particles~\cite{Netz97JphysI,Park96JPhysI,Dommersnes99EPJB,Dommersnes02BiophysJ,Bitbol10PRERC} and dynamical behaviors such as mobility and diffusion~\cite{CortezSJSC01,LevinePRE01,PeskinAN02,AtzbergerJCP07,CamleyPRE11,CamleySM13}. Either sharp cutoffs are used, with excellent approximate results~\cite{Park96JPhysI,Dommersnes99EPJB,Dommersnes02BiophysJ,Bitbol10PRERC,LevinePRE01}, or smooth cutoffs in numerical works that have the effect of distributing the applied forces over small finite regions~\cite{CortezSJSC01,PeskinAN02,AtzbergerJCP07,CamleyPRE11,CamleySM13}.

Owing to the simplicity of the pointlike approach, it is possible to go beyond the SD problem and to provide analytical or semi-analytical results while taking into account several complications, such as (i) the bilayer character of the membrane~\cite{CamleySM13,SekiPRE14} and (ii) the spontaneous curvature of the inclusion~\cite{NajiPRL09,Quemeneur14PNAS,Morris15PRL}.

A real membrane is not simply a 2D viscous slab, it is made up of two contacting fluid monolayers (labelled as $\pm$), each with its own 2D viscosity $\eta_2^\pm$. The question of the continuity, or discontinuity, of the lipids velocity across the separation between the two monolayers is important~\cite{MerkelJPhys89,Evans94CPL,Seifert93EPL}.
Whereas it is legitimate to impose a no-slip boundary condition at the interface between each monolayer and its contacting solvent, and  at the interface between an inclusion's boundary and its contacting monolayer, it is in general necessary to allow for some velocity discontinuity $\Delta\bm v=\bm v^+ -\bm v^-$ at the interface between the two monolayers~\cite{Seifert93EPL}. Intermonolayer sliding occurs essentially because there is very little interdigitation of the lipids tails at the interface between the two monolayers. The relevant parameter is the intermonolayer friction coefficient $b$, which plays the role of a discrete viscosity: the stress transmitted through the interface is $b\Delta\bm v$~\cite{MerkelJPhys89}, very much like the viscous stress $\eta\partial V_x/\partial z$ in a bulk fluid. The larger $b$, the more continuity is imposed, and the smaller $b$, the more sliding is allowed. In practice, for membranes, $b$ has been reported over a wide range $\SIrange{e6}{e9}{J.s.m^{-4}}$~\cite{OtterBJ07,FournierPRL09}, so we shall leave open the possibility of intermonolayer sliding.

Here, we study two situations in which the bilayer structure is relevant. We first consider an inclusion that is embedded in only one of the two monolayers, or adhering to one of them, as it is the case for the BAR family of proteins~\cite{SimunovicTCB15}. Note that membrane curvature effects~\cite{Quemeneur14PNAS} and the possible coupling with other order parameters~\cite{NajiBJ07,CamleyPRE12,NajiPRL09} are disregarded. When a force is applied to the object, it sets into motion the fluid of the monolayer in which it stands, but it is the intermonolayer friction that drags the fluid of the other monolayer. We therefore expect a modified SD law in which intermonolayer friction plays a role. Indeed, we find a corrective factor $\sqrt{1+\eta_2/(4ba^2)}$ in the logarithm of  the SD law, which can be important or not depending on the value of $b$.
Then, we consider an inclusion that spans the whole bilayer, assuming however different monolayer viscosities $\eta_2^+$ and $\eta_2^-$. We find that except for extremely small (unphysical) values of the intermonolayer friction $b$, the velocity is almost perfectly the same in the two monolayers and the SD law holds upon changing $\eta_2$ into $\eta_2^+ + \eta_2^-$ (note that the 2D viscosity of a thin layer is  proportional to its thickness). So, even if there is a strong discontinuity between the monolayers viscosities, everything happens as if the bilayer was a homogeneous fluid with the average viscosity.

Next, we consider a curvature-inducing protein, i.e., a particule that promotes a local curvature of the membrane. To our knowledge, previous papers did not attempt to solve consistently the coupled dynamical equations for the 2D membrane flow and the 3D solvent flow~\cite{NajiPRL09,Quemeneur14PNAS,Morris15PRL}. Indeed, in Refs.~\cite{NajiPRL09,Quemeneur14PNAS} the authors directly add the SD dissipation to an extra dissipation term calculated from the membrane  dynamics, while in Ref.~\cite{Morris15PRL}, although a nonlinear flow-curvature coupling is taken into account, the authors impose an asymptotic matching of their solution to the SD solution. Note that they also put forward a tension-induced deformation of the protein, which they claim to be responsible for the observations of Ref.~\cite{Quemeneur14PNAS}. Our pointlike approximation allows for a complete treatment of the 2D-3D coupled problem in the quasi-planar geometry (neglecting protein deformations). We take into account, as in Refs.~\cite{Quemeneur14PNAS,Demery10PRL}, the response of the membrane deformation to the applied force (this effect was neglected in Refs.~\cite{NajiPRL09,Morris15PRL}) and we show, in the quasi-planar approximation, that the total friction is the sum of the SD friction and that due to the pull-back created by the membrane deformation.

\section{Saffman-Delbr\"uck}
\label{sec:SD}

Before embarking on the SD problem, let us go back to the 3D Stokes law and the Stokes paradox. When a pointlike force $\bm f$ is applied at the origin of a 3D fluid, the velocity field is given (in the limit of low Reynolds numbers) by the Oseen tensor~\cite{DoiEdwards_book}. In reciprocal space, this reads $\bm{v}(\bm{q})=(\eta q^2)^{-1}(\mathbf{I}-\hat{\bm q}\otimes\hat{\bm q})\cdot\bm{f}$, where $\mathbf{I}$ is the identity tensor and $\hat{\bm q}$ the unit vector parallel to $\bm{q}$. Assuming that the force is applied to a particle that transmits it to the fluid, one can calculate the velocity $v_p$ of the particle by looking at the velocity of the fluid at the origin. Discarding angular factors, the latter is given by $v_p/f\propto\int {\rm{d}}^dq/(\eta q^2)$, with $d$ the space dimension. In 3D, this integral converges at low $q$ but diverges at high $q$, so a cutoff is necessary to regularize it. Taking $q_\mathrm{max}=1/a$, where $a$ is the size of the particle, we obtain $v_p/f=1/(3\pi^2\eta a)$. We recover the Stokes law except for an incorrect multiplicative factor which can be viewed as an imprecision on the radius of the particle. In 2D, however, the integral diverges at low $q$ and the mobility is  found to be infinite, in agreement with the Stokes paradox. In the case of membranes, it is the solvent's viscosity that provides a regularization~\cite{SaffmanPNAS75}.

\begin{figure}
\centerline{\includegraphics[width=.8\columnwidth]{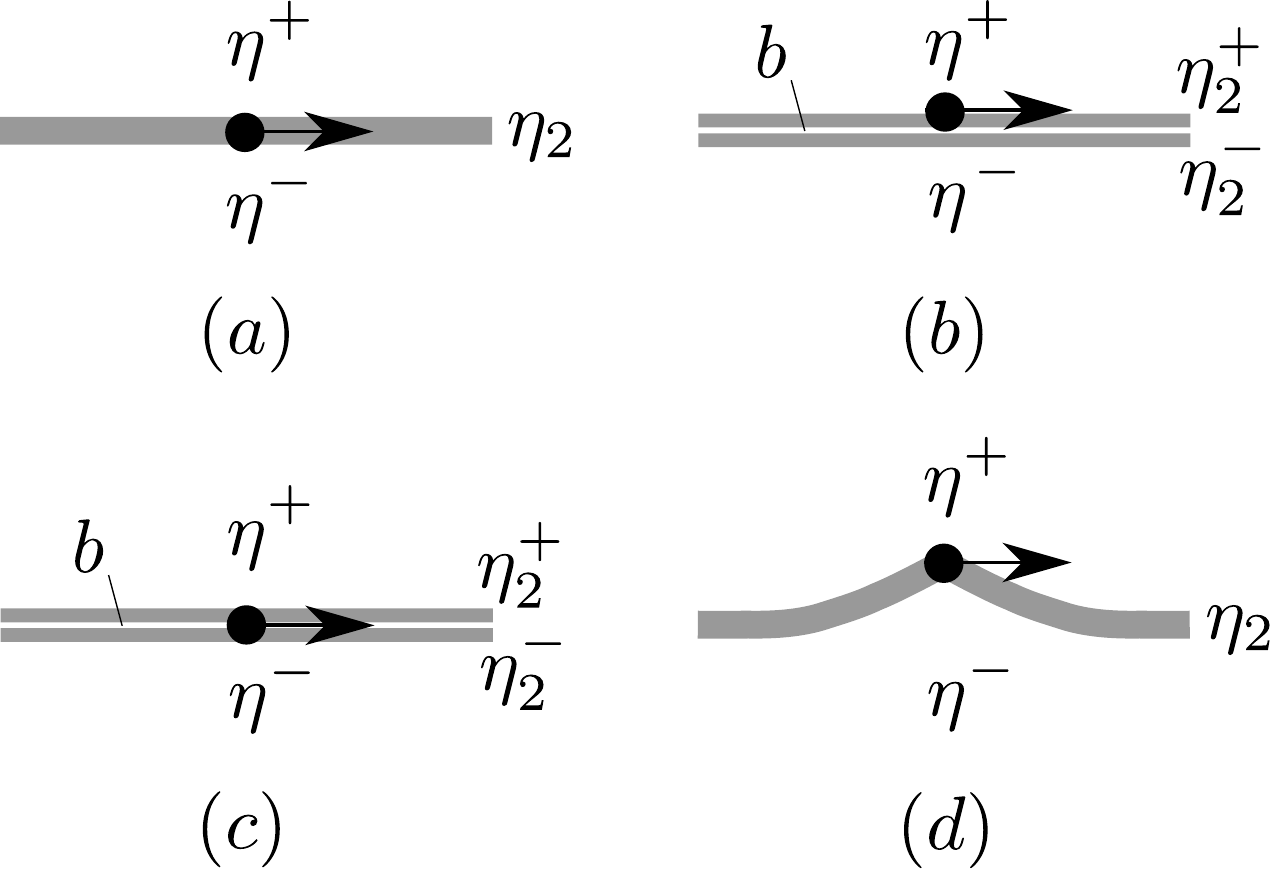}}
\caption{Sketch of an inclusion pulled in a membrane embedded in a solvent with 3D viscosities $\eta^+$ above and $\eta^-$ below. In (a) and (d) the membrane is a homogeneous medium with 2D viscosity $\eta_2$ while in (b) and (c) it consists of two adjacent media (monolayers) of 2D viscosities $\eta_2^+$ and $\eta_2^-$, subject to an intermonolayer friction with coefficient $b$. In (b) the inclusion is in the upper monolayer only while in the other pictures it spans the whole membrane (or bilayer). In (d) the inclusion creates a local curvature that relaxes away.}
\label{schema}
\end{figure}

In the Saffman-Delbr\"uck problem, a particle of position $\bm R(t)$ is dragged by a constant force $\bm f$ in a fluid membrane embedded in a solvent (Fig.~\ref{schema}a). The membrane is treated as a structureless 2D fluid of viscosity $\eta_2$ in the $z=0$ plane, and inertia is neglected. The membrane lies in a bulk solvent of viscosities $\eta^\pm$ in the two half spaces (indicated with the superscript $\epsilon=\pm$). Here, contrary to the traditional approach~\cite{SaffmanJFM76} we are going to treat the particle as pointlike, which will greatly simplify the calculations. This entails an approximation, hence validity conditions that we shall discuss in detail at the end of Sec.~\ref{sec:SDmomo}. Calling $\bm{v}$ the velocity field in the membrane and $\bm{V}^\pm$ the velocity fields in the bulk, the equations describing the problem are
\begin{eqnarray}
&&\eta^\pm\nabla^2\bm V^\pm - \bm\nabla P^\pm=0,
\\
&&\eta_2\bar{\nabla}^2\bm v - \!\bar{\bm{\nabla}}p 
+\bm\sigma^+ + \bm\sigma^-
+\bm f\delta(\bm r-\bm R)=0,
\label{membeq}
\\
&&\dot{\bm R}=\bm v(\bm R),
\label{dotR}
\end{eqnarray}
where
\begin{equation}
\bm\sigma^\pm=\pm\eta^\pm\!\left.(\partial_z\bar{\bm V}^\pm+\bar{\bm\nabla}V_z^\pm)\right|_{z=0}
\end{equation}
are the tangent viscous stresses
transmitted to the membrane by the bulk flow. The first equation is the 3D Stokes equation describing the flow in the solvent, the second equation is the Stokes equation in the membrane,
and the third equation is a no-slip  condition on the pointlike particle reflecting its transport by the membrane flow. Note that since inertia is neglected, the force $\bm f$ applied to the particle is directly transmitted to the 2D membrane fluid. Here,
 $p$ and $P^\pm$ are the excess pressure fields (pressure minus the pressure at infinity), the dot denotes time derivative and the bar denotes the projection onto the $(x,y)$ plane of 3D vectors ($\bar{\bm\nabla}=\bm e_x\partial_x+\bm e_y\partial_y$ and $\bar{\bm V}=V_x\bm e_x+V_y\bm e_y$).  These equations are supplemented by the incompressibility conditions $\bm \nabla\cdot\bm V^\pm =0$ and $\bm {\bar\nabla}\cdot\bm v =0$, and by the no-slip and continuity conditions $\bar{\bm V}^\pm|_{z=0}=\bm v$ and $V^\pm_z|_{z=0}=0$.

As we are dealing with a pointlike force, solving for the membrane flow is  equivalent to determining the Oseen-like tensor in this geometry~\cite{LubenskyPF96,Oppenheimer10PRE}. We start by eliminating the bulk variables. For this, we Fourier transform in the $(x,y)$ plane while keeping  $z$  in real space. Let $\bm{V}^\pm(\bm{q},z)=\int {\rm d}^2r\,\bm V^\pm(\bm r,z)e^{-i\bm q\cdot\bm r}$. We decompose it as $\bm V^\pm=V^\pm_\parallel\hat{\bm{q}}+V^\pm_\perp\hat{\bm{q}}_\perp+V^\pm_z\hat{\bm e}_z$, where $\hat{\bm{q}}=\bm q/q$ and $\hat{\bm{q}}_\perp=\hat{\bm{e}}_z\times\hat{\bm{q}}$, and likewise $\bm v=v_\parallel\hat{\bm{q}}+v_\perp\hat{\bm{q}}_\perp$. Incompressibility yields $ v_\parallel(\bm{q})=0$.
The bulk equations read $\eta^\pm(\partial^2_z-q^2)(V_\parallel^\pm,V_\perp^\pm,V_z^\pm)=(iqP^\pm,0,\partial_zP^\pm)$ and $\partial_zV_z^\pm+iqV_\parallel^\pm=0$.
Solving them with the boundary condition $\bm V^\pm(\bm q,z)|_{z=0}=v_\perp(\bm q)\hat{\bm{q}}_\perp$ yields $P^\pm(\bm{q},z)=0$ and $\bm V^\pm(\bm q,z)=v_\perp(\bm q)\exp(\mp qz)\hat{\bm q}_\perp$. It follows that $\bm\sigma^\pm=-\eta^\pm q v_\perp(\bm q) \hat{\bm{q}}_\perp$~\cite{Seifert93EPL}, and the hydrodynamic equations in the membrane reduce simply to
\begin{eqnarray}
&&-\eta_2 q^2 v_\perp -2\eta q v_\perp + \bm f e^{-i\bm q\cdot\bm R}\cdot\hat{\bm q}_\perp =0, 
\\
&&v_\parallel=0,
\end{eqnarray}
where $2\eta=\eta^++\eta^-$.
The solution for $\bm v(\bm q)$ is then
\begin{equation}
\bm v(\bm q)  = \mathbf O(\bm q)\cdot\bm f e^{-i\bm q\cdot\bm R},
\quad
\mathbf O(\bm q)=\frac{\mathbf{I}-\hat{\bm q}\otimes\hat{\bm q}}{2\eta q + \eta_2 q^2},
\label{modifiedoseen}
\end{equation}
with $\mathbf O$ the Oseen-like tensor in the SD geometry~\cite{LubenskyPF96,Oppenheimer10PRE}. Assuming then $\bm f=f\bm e_x$, we obtain the particle's  velocity as
\begin{equation}
v_p=\bm e_x\cdot\bm v(\bm R)=\int
\frac{q\,{\rm d}q\,{\rm d}\theta}{(2\pi)^2}
\frac{1-\hat q_x^2}{2\eta q + \eta_2 q^2}f,
\end{equation}
where $\hat q_x=\cos\theta$.
However, due to its ultraviolet (high-$q$) behavior, this integral diverges logarithmically. Since the short-scale velocity gradients of $v(\bm r)$ are located near $\bm r=\bm R$ while in reality the particle has a uniform velocity field for $r<a$ (it is a solid), we may resolve this problem by eliminating the Fourier modes with wavevectors larger than the inverse particle radius $a^{-1}$ (see a similar approach in the Appendix B of Ref.~\cite{CamleySM13}). To simplify, we use a sharp cutoff $q_\mathrm{max}= a^{-1}$,
and integrating over $q$ in the range $[0,a^{-1}]$ yields
\begin{equation}
v_p=\frac{f}{4\pi\eta_2}\ln\left(1+\frac{\ell}{a}\right),
\label{SDlaw}
\end{equation}
with the SD length
\begin{equation}
\ell=\frac{\eta_2}{\eta^+ + \eta^-}.
\end{equation}
We shall discuss more precisely in Sec.~\ref{sec:SDmomo} the validity conditions of our pointlike method. Let us simply note here that we need to assume $a\ll\ell$ otherwise we would be neglecting the Fourier modes at the scale of the SD length, which are physically important.
The condition $a\ll\ell$ is also a condition of validity of the original SD law~\cite{SaffmanPNAS75,SaffmanJFM76,HughesJFM81}. Note that it is very well satisfied for proteins in membranes, since $a$ lies in the nanometer range while $\ell$ lies in the micron range. 

In this limit we thus obtain $v_p\simeq f(4\pi\eta_2)^{-1}\ln(\ell/a)$. Therefore the particle's mobility $\tilde\mu=v_p/f$ is given, within our approximation scheme, by
\begin{equation}
\label{SDlaw2}
\tilde\mu=\frac{1}{4\pi\eta_2}\ln\frac\ell a.
\end{equation}
This is the SD law, except for an extra factor of order unity multiplying the particle radius, namely $2/e^\gamma\simeq1.1$. Because this factor is within the logarithm, the  prefactor obtained here, i.e., $(4\pi\eta_2)^{-1}$, is exactly that of the SD law (contrary to the 3D case in which the radius appears in a the Stokes power law). Numerically, with the typical parameters $a\simeq\SI{3}{nm}$, $\eta^\pm\simeq\SI{e-3}{J.s.m^{-3}}$ and $\eta_2\simeq\SI{e-9}{J.s.m^{-2}}$, we find that $\mu$ and $\tilde\mu$ differ only by $2\%$. Note that while it is formally important to have an exact result for a perfect disc (the SD law), real objects embedded in membranes, like integral proteins, are not perfect cylinders, but rather cylindrical-like or conical-like inclusions with an inhomogeneous radius, so that uncertainties on the radius are actually not so important.

\section{Saffman-Delbr\"uck for an inclusion in one monolayer}
\label{sec:SDmomo}

The diffusion behavior of solid particles embedded in a single monolayer of a bilayer membrane was studied numerically in Ref.~\cite{CamleySM13}, and discussed analytically in Ref.~\cite{Hill14PRCA} for supported membranes using a phenomenological friction/slip description. The related problem of the diffusion of liquid domains within a monolayer (thus involving lipid flow inside the domains) was discussed in Ref.~\cite{SekiPRE14}.

We  consider an inclusion embedded in the upper monolayer of a membrane, or simply adhering to it (Fig.~\ref{schema}b). 
The membrane is treated as a bilayer with monolayer viscosities $\eta_2^\pm$ and intermonolayer friction $b$, embedded in a bulk fluid with viscosities $\eta^\pm$. The force $\bm f^+$ applied to the particle is transmitted to the upper monolayer, so that the dynamical equations become
\begin{eqnarray}
&&\eta^\pm\nabla^2\bm V^\pm - \bm\nabla P^\pm=0,
\\
&&\eta_2^+\bar\nabla^2\bm v^+ 
-\bar{\bm\nabla}p^+
+ \bm\sigma^+
-b\Delta\bm v
+ \bm f^+\delta(\bm r-\bm R)=0,\hspace{16pt}
\\
&&\eta_2^-\bar\nabla^2\bm v^- 
-\bar{\bm\nabla}p^-
+ \bm\sigma^-
+ b\Delta\bm v = 0,
\\
&&\dot{\bm R}=\bm v^+(\bm R),
\label{velplus}
\end{eqnarray}
where $\Delta\bm v=\bm v^+-\bm v^-$. These equations are supplemented by the following incompressibility and continuity equations: $\bm \nabla\cdot\bm V^\pm =0$, $\bm {\bar\nabla}\cdot\bm v^\pm =0$, $\bar{\bm V}^\pm|_{z=0}=\bm v^\pm$ and $V^\pm_z|_{z=0}=0$. With respect to the previous problem, eq.~(\ref{membeq}) has been splitted into two equations (one for each monolayer), intermonolayer friction has been added, and the no-slip condition~(\ref{velplus}) expressing the transport of the particle involves now only the upper monolayer flow.
Going to Fourier space and eliminating the bulk variables as previously yields the membrane equations:
\begin{eqnarray}
&&- \eta_2^+ q^2 v_\perp^+
- \eta^+ q v_\perp^+
-b\Delta v_\perp
+ \bm f^+e^{-i\bm q\cdot\bm R}\cdot\hat{\bm q}_\perp = 0,
\hspace{20pt}
\\
&&- \eta_2^- q^2 v_\perp^-
- \eta^- q v_\perp^-
+ b\Delta v_\perp = 0,
\\
&& v_\parallel^\pm = 0.
\end{eqnarray} 
Solving for the monolayer velocities, we  obtain $\bm v^\pm(\bm q)=\mathbf O^\pm_+(\bm q) \cdot \bm f^+e^{-i\bm q\cdot\bm R}$, with
\begin{eqnarray}
&&\mathbf O_+^\epsilon(\bm q) = A_+^\epsilon(q) \left(\mathbf{I}-\hat{\bm q}\otimes\hat{\bm q}\right),
\label{Oseen2}
\\
&&A_+^\epsilon(q)=
\frac{b+q(\eta^- + \eta_2^-q)\delta_{\epsilon,+}}{q D(q)},
\end{eqnarray}
where
\begin{eqnarray}
D(q)&&=b\left[\eta^+ + \eta^- + (\eta_2^+ + \eta_2^-)q\right]
\nonumber\\
&&+\, q(\eta^+ + \eta_2^+ q)(\eta^- + \eta_2^- q).\hspace{18pt}
\end{eqnarray}
The tensor $\mathbf O_-^\pm$ giving the velocities for a pointlike force $\bm f^-$ applied to the lower monolayer is obtained by exchanging the $+$ and $-$ signs.
These Oseen-like tensors were first derived in Ref.~\cite{CamleySM13}.

In order to calculate, within our regularized pointlike approximation, the mobility $\tilde\mu_m$ of a particle embedded in the upper monolayer, we take $\bm f^+=f^+\bm e_x$, which yields the particle velocity
$v_p=\bm e_x\cdot\bm v^+(\bm R)=\tilde\mu_m f^+$, with
\begin{equation}
\label{mu_tilde_m_num}
\tilde\mu_m=\int_0^{a^{-1}}\!\!\!\frac{q\,{\rm d}q\,{\rm d}\theta}{(2\pi)^2}\,A_+^+(q)\sin^2\theta.
\end{equation}
As previously, we have regularized the integral by using an upper wavevector cutoff equal to the inverse of the particle's radius.

We first consider the symmetric situation where $\eta^+=\eta^-=\eta^\pm$ and $\eta_2^+=\eta_2^-=\eta_2^\pm$. In this case the solution is analytic, given exactly by
\begin{eqnarray}
\tilde\mu_m&=&\frac1{8\pi\eta_2^\pm}
\ln\left[
\left(1+
\frac{\eta_2^\pm}{\eta^\pm a}
\right)
\sqrt{1+\frac{\eta_2^\pm+\eta^\pm a}{2a^2b}}
\,\,\right]
\nonumber\\
&+&\frac
{\eta\ln\!\left(\frac{4ab+\eta^\pm-\sqrt{{\eta^\pm}^2-8b\eta_2^\pm}}
{4ab+\eta^\pm+\sqrt{{\eta^\pm}^2-8b\eta_2^\pm}}\right)}
{16\pi\eta_2^\pm\sqrt{{\eta^\pm}^2-8b\eta_2^\pm}}.
\label{mu12}
\end{eqnarray}
Assuming $a\ll\ell=\eta_2^\pm/\eta^\pm$ as in the previous section, the first term $\tilde\mu^{(1)}$ of $\tilde\mu$, which is also the term dominant at large $b$, can be simplified into
\begin{equation}
\label{mu_m_racine}
\tilde\mu_m^{(1)}\simeq
\frac1{8\pi\eta_2^\pm}
\ln\left(
\frac{\ell}{a}
\sqrt{1+\frac{\eta_2^\pm}{2a^2b}}
\right).
\end{equation}

\begin{figure}
\centerline{\includegraphics[width=.9\columnwidth]{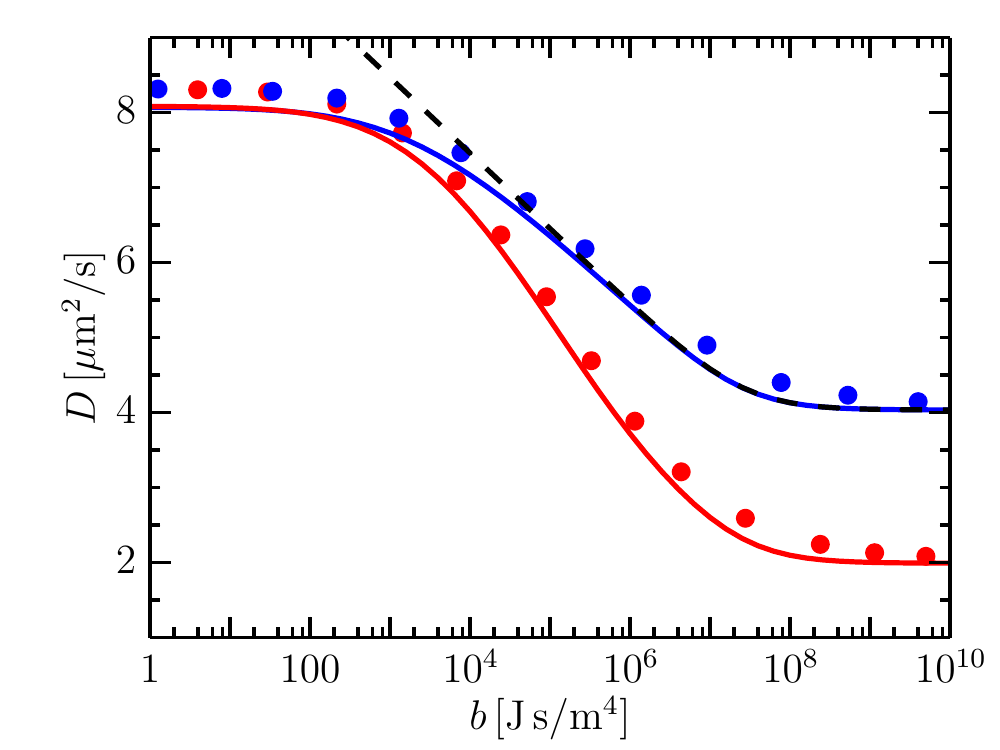}}
\caption{Diffusion coefficient $D=k_\mathrm{B}T\tilde\mu_m$, as a function of the intermonolayer friction $b$, for a protein in the upper monolayer of a membrane (i) in an infinite solvent, i.e., $H=\infty$ (blue line) and (ii) in a supported membrane at a distance $H$ from a substrate (red line). Dashed black line: approximation $D\simeq k_\mathrm{B}T\tilde\mu_m^{(1)}$, valid for large $b$ in an infinite solvent. The parameters are identical to those of Camley and Brown in Ref.~\cite{CamleySM13}, Fig.~7, i.e., $T=\SI{319}{K}$, $\eta^\pm=\SI{e-3}{J.s.m^{-3}}$, $\eta_2^\pm=\SI{2e-10}{J.s.m^{-2}}$, $a=\SI{2}{nm}$ and $H=\SI{1}{nm}$. The red and blue data points were extracted from the Fig.~7 of Ref.~\cite{CamleySM13}. Our theory fits well the numerical data with no adjustable parameter.}
\label{Dvsb}
\end{figure}

The mobility is related to the diffusion coefficient by  Einstein's relation $D=k_\mathrm{B}T\tilde\mu_m$. The blue curve of Fig.~\ref{Dvsb} shows $D$ versus the intermonolayer friction $b$ for infinite solvents, as obtained from eq.~(\ref{mu12}). We see that $\tilde\mu_m\simeq\tilde\mu_m^{(1)}$ for large $b$, as evidenced by the black dashed line in Fig.~\ref{Dvsb}. Since the usual physical range of $b$ for lipid membranes lies in this region, we infer that in practice
\begin{equation}
\tilde\mu_m\simeq
\tilde\mu_m^{(1)}.
\end{equation}
This formula is typically valid for $b$ larger than $\SI{e5}{J.s.m^{-4}}$.

We deduce that the SD law is influenced by $b$ when $b<\eta_2^\pm/(2a^2)$.
The limit $b\to\infty$ gives exactly the SD law. Indeed, $\tilde\mu_m\to\tilde\mu$ with $\eta_2$ replaced by $2\eta_2^\pm$. This comes from the fact that both monolayers are fully dragged by the applied force. The limit $b\to0$ gives  $\tilde\mu_m\to2\tilde\mu$, as apparent in Fig.~\ref{Dvsb}, a result already pointed out in Refs.~\cite{CamleySM13,SekiPRE14}. This stems from the fact that only the monolayer containing the inclusion is dragged by the force. These behaviors are easily deduced from the form of $\mathbf{O}^+_+(q)$, which converges to $\mathbf{O}(q)$ with $\eta_2=2\eta_2^\pm$ when $b\to\infty$ and to $\mathbf{O}(q)$ with  $\eta_2=\eta_2^\pm$ when $b\to0$.

In Fig.~\ref{Dvsb}, we plotted together with our analytical curves several data points extracted from the numerical calculations of Camley and Brown~\cite{CamleySM13} who addressed the same problem. Note the fair agreement, with no adjustable parameter.
The presence of a substrate at a distance $H$ below the membrane can be taken into account very simply within our model by replacing $\eta^-$ by $\eta^-\coth(q H)$ in the Oseen-like tensor (\ref{Oseen2})~\cite{LubenskyPF96,Oppenheimer10PRE}. Then the numerical integral (\ref{mu_tilde_m_num}) gives the red curve in Fig.~\ref{Dvsb} which displays also a fair agreement with the numerical results of Ref.~\cite{CamleySM13}.

In the general asymmetric situation, such that $\eta^+\ne\eta^-$ or $\eta_2^+\ne\eta_2^-$, the integral (\ref{mu_tilde_m_num}) giving $\tilde\mu_m$ must be done numerically, because the roots of $D(q)$ are complicated.
It is possible, however, to get analytical results in the following two situations:

(i) For $b\to\infty$, $\mathbf{O}^+_+(q)\to\mathbf{O}(q)$ with $\eta_2$ replaced by $\eta_2^++\eta_2^-$. Therefore the mobility $\tilde\mu_m$ tends to $\tilde\mu_\infty$ with
\begin{equation}
\tilde\mu_\infty=\frac1{4\pi(\eta_2^+ + \eta_2^-)}
\ln\frac{\eta_2^+ + \eta_2^-}{(\eta^+ + \eta^-)a}.
\end{equation}
We recover the SD law. Everything happens as if the particle were embedded in a single layer with a 2D viscosity equal to the sum of those of the monolayers (recall that the 2D viscosity of a thin layer is proportional to its thickness). For ordinary values of the viscosities, i.e., $\eta^\pm\simeq\SI{e-3}{J.s.m^{-3}}$ and $\eta_2^\pm\simeq\SI{e-9}{J.s.m^{-2}}$, $\tilde\mu_m$ is well approximated by $\tilde\mu_\infty$ as soon as $b\so\SI{e8}{J.s.m^{-4}}$ (like in Fig.~\ref{Dvsb}).

(ii) In the somewhat formal proportional case $\eta^-/\eta^+=\eta_2^-/\eta_2^+=\alpha$, we obtain analytically the following generalization of eq.~(\ref{mu_m_racine}):
\begin{equation}
\label{mobilitymono}
\tilde\mu_m\simeq
\frac{1}{4\pi(1+\alpha)\eta_2^+}\ln\left[
\frac{\ell}{a}\left(
1 + \frac{2\alpha}{1+\alpha}\,\frac{\eta_2^+}{2a^2b}
\right)^{\alpha/ 2}
\right],
\end{equation}
valid also typically for $b\so\SI{e5}{J.s.m^{-4}}$, like in Fig.~\ref{Dvsb}.

Let us now discuss in more detail the conditions of validity of our pointlike approximation. We do eliminate all the Fourier modes with wavevectors larger than the inverse particle radius $a^{-1}$. Obviously, we must not eliminate modes having a physical meaning (stemming from relevant characteristic lengths) and contributing significantly to the integral of the Oseen-like tensor. First, let us note that this will never significantly be the case for transmembrane proteins, because $a$ is is in the nanometer range and Fourier modes with smaller wavelenghts are unphysical (they are of the order of the membrane thickness or of several lipid widths). In other words, there is already an implicit cutoff in the nanometer range in the system. So, whatever the characteristic lengths involved in the Oseen-like tensor, our approximate method can be safely applied to membrane proteins.
Now, if we were to apply our method to somewhat larger particles, e.g., liquid domains, it would be necessary to investigate whether the integral of the modes between $a^{-1}$ and the inverse nanometer range contribute negligibly to the total integral or not. For instance, our method would fail for solid particles larger than the SD length since it would yield a mobility different from that calculated by Hughes et al.~\cite{HughesJFM81}.

\section{Saffman-Delbr\"uck for an inclusion spanning the bilayer}

\begin{figure}
\centerline{\includegraphics[width=.9\columnwidth]{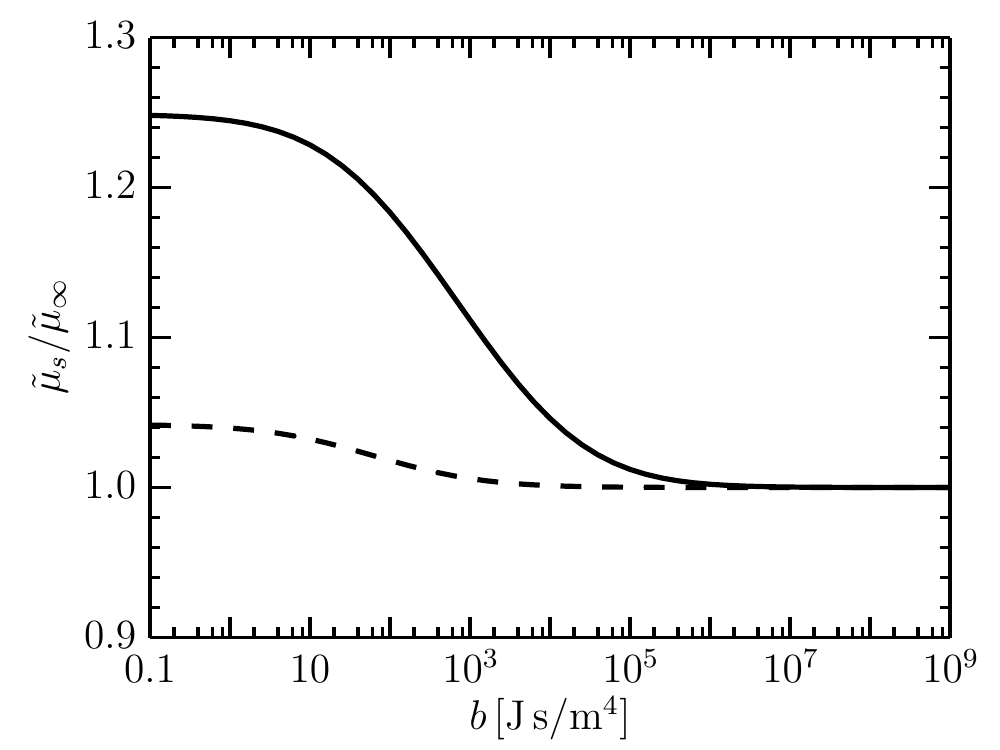}}
\caption{
Mobility $\tilde\mu_s$ of a particle with radius $a=\SI{3}{nm}$ spanning an asymmetric bilayer as a function of the intermonolayer friction $b$, normalized by its limit $\tilde\mu_\infty$ for $b\to\infty$. The viscosities are $\eta^-=10\eta^+=\SI{e-2}{J.s.m^{-3}}$, $\eta_2^+=10\eta_2^-=\SI{e-8}{J.s.m^{-2}}$ (solid curve) and $\eta^+=\eta^-=\SI{e-3}{J.s.m^{-3}}$, $\eta_2^+=10\eta_2^-=\SI{e-8}{J.s.m^{-2}}$ (dashed curve).
}
\label{bib}
\end{figure}

We now apply our method in order to calculate the mobility $\tilde\mu_s$ of an inclusion spanning the whole membrane, while taking into account the bilayer structure of the latter. The force $\bm f$ applied to the inclusion is now transmitted to both monolayers in the form of two pointlike forces $\bm f^\pm$.
If $\eta_2^+\ne\eta_2^-$ (or $\eta^+\ne\eta^-$) we expect, by lack of symmetry, these forces to be different. They are determined by the conservation of the total force and by the no-slip boundary condition at the surface of the particle:
\begin{eqnarray}
\label{s1}
&&\bm f^+ + \bm f^- = \bm f,
\\
\label{s2}
&&\bm v^+(\bm R) = \bm v^-(\bm R)=\dot{\bm R}.
\label{noculotte}
\end{eqnarray}
Using eq.~(\ref{Oseen2}) and the linearity of the problem, we get
\begin{eqnarray}
\bm v^\pm(\bm q)&=&
\mathbf O^\pm_+\cdot\bm f^+e^{-i\bm q\cdot\bm R} + \mathbf O^\pm_-\cdot\bm f^-e^{-i\bm q\cdot\bm R}
\nonumber\\
&=&\frac{\mathbf{I}-\hat{\bm q}\otimes\hat{\bm q}}{q D(q)}\cdot
\left[
b\bm f
+ q(\eta^\mp + \eta_2^\mp q)\bm f^\pm
\right]e^{-i\bm q\cdot\bm R}.\hspace{12pt}
\end{eqnarray}
Using our regularized pointlike approximation, we obtain $\bm v^\pm(\bm R)=b I_0 \bm f + (\eta^\mp I_1 + \eta_2^\mp I_2) \bm f^\pm$, with
\begin{equation}
I_n=\int_0^{a^{-1}}\!\!\mathrm{d}q\,\frac{q^n}{4\pi D(q)}.
\end{equation}
Solving then eqs.~(\ref{s1})--(\ref{s2}) gives 
\begin{equation}
\bm f^\pm =
\frac{\eta^\pm I_1 + \eta_2^\pm I_2}
{(\eta^++\eta^-) I_1 + (\eta_2^+ + \eta_2^-)I_2}\bm f,
\end{equation}
and then $\bm v_p=\tilde\mu_s \bm f$, with
\begin{equation}
\tilde\mu_s=
b I_0
+ \frac{(\eta^- I_1 + \eta_2^- I_2)(\eta^+ I_1 + \eta_2^+ I_2)}
{(\eta^++\eta^-) I_1 + (\eta_2^+ + \eta_2^-)I_2}.
\label{mu_tilde_s}
\end{equation}
Note that in the completely symmetric case $\eta^+=\eta^-=\eta$ and $\eta_2^+=\eta_2^-=\frac12\eta_2$ the dependence in $b$ disappears and $\tilde\mu_s$ reduces to $\tilde\mu$.
The mobility $\tilde\mu_s$ must be studied numerically~\cite{Mathematica} because $I_n$ has no simple analytical form. We find that  $\tilde\mu_s$ increases slightly as $b$ decreases, in a way that is enhanced by the asymmetry of the viscosities (fig.~\ref{bib}). However, this effect is actually rather negligible, since for ordinary viscosities (as in fig.~\ref{bib}) it requires $b\io\SI{e4}{J.s.m^{-4}}$, well below any experimental value. We may therefore take the limit $b\to\infty$, which yields $D(q)\simeq b[\eta^+ + \eta^- + (\eta_2^+ + \eta_2^-)q]$, $I_n\propto1/b$, and thus
\begin{equation}
\tilde\mu_s\simeq bI_0\to\tilde\mu_\infty.
\end{equation}
We recover again the SD
law with $\eta_2$  replaced by $\eta_2^+ + \eta_2^-$.
Physically, intermonolayer friction can be disregarded for particles spanning the bilayer, because monolayer slippage nearby the particle is forbidden by the no-slip conditions~(\ref{noculotte}) at the particle's boundary.

\section{Saffman-Delbr\"uck for an inclusion curving the membrane}

Let us finally apply our method to membrane inclusions that curve the membrane. Such particles are usually either transmembrane proteins with a conical shape that bind the surrounding lipids, thus imposing a local curvature to the membrane~\cite{Gruler75ZN,Leibler86JP,Goulian93EPL}, or nonflat capping  proteins adhering to the membrane~\cite{Prevost15NC}, with the same result. Experiments have been made also with larger adhering beads~\cite{VanderWel16SR}.

We will confine ourselves to weakly deformed membranes, described by their elevation $z=h(\bm r)$ above the reference plane $\bm r=(x,y)$.
The elastic energy of the system consisting of the membrane and the inclusion can  be expressed as
\begin{equation}
\label{H}
\mathcal{H}=\int d^2r\left[\frac\kappa2(\nabla^2h)^2+\frac\sigma2(\bm\nabla h)^2
+B\,G(\bm r-\bm R)\nabla^2h\right].
\end{equation} 
The first two terms correspond to the Helfrich Hamiltonian and describe the bending energy of the membrane and the energy associated with its tension $\sigma$~\cite{Helfrich73}. The third term models an isotropic inclusion located at the in-plane position $\bm R$ that promotes membrane curvature with a strength $B$, in the way of Ref.~\cite{Quemeneur14PNAS}. The function $G(r)$ is a generic function describing the envelope of the protein influence over the membrane, e.g., a Gaussian with a width comparable to the protein's radius~\cite{NajiPRL09,Quemeneur14PNAS}. Note that in this model the actual curvature set by the inclusion depends on the elastic response of the membrane.

Assuming that the flow within the membrane remains quasi 2D, which is standard in the limit of small deformations~\cite{Brochard75JPhys,Seifert93EPL}, and disregarding the membrane bilayer structure for the sake of simplicity, the dynamical equations of the system can be written as
\begin{eqnarray}
&&\eta^\pm\nabla^2\bm V^\pm - \bm\nabla P^\pm=0,
\\
&&\eta_2\bar{\nabla}^2\bm v - \!\bar{\bm{\nabla}}p 
+\bm\sigma^+ + \bm\sigma^-
+\left(\bm f-\frac{\partial\mathcal{H}}{\partial\bm R}\right)\delta(\bm r\bm- R)=0,\hspace{18pt}
\label{eqtan}
\\
&&-\frac{\delta\mathcal{H}}{\delta h}+\Sigma^++\Sigma^-=0,
\label{eqnor}
\\
&&\dot{\bm R}=\bm v(\bm R).
\end{eqnarray}
where
\begin{equation}
\Sigma^\pm=\pm(2\eta^\pm\partial_zV_z^\pm-P^\pm)|_{z=0}.
\end{equation}
The first equation is the bulk Stokes equation. The second equation is the Stokes equation for the membrane planar flow, including the force density transmitted by the particle. The third equation is the balance of the stresses normal to the membrane, with $\Sigma^\pm$ the stresses transmitted by the bulk. The last equation is the no-slip condition expressing the transport of the particle. These equations must be supplemented by the incompressibility relations $\bm \nabla\cdot\bm V^\pm =0$ and $\bm {\bar\nabla}\cdot\bm v =0$, and by the continuity conditions $\bar{\bm V}^\pm|_{z=0}=\bm v$ and $V^\pm_z|_{z=0}=\dot h$.

Let us now express them in mixed reciprocal-direct space, as in Sec.~\ref{sec:SD}, so as to eliminate the bulk velocities~\cite{Seifert93EPL}. The boundary conditions read $V^\pm_z(\bm q,z)|_{z=0}=\dot h(q)$ and $\bar{\bm V}^\pm(\bm q,z)|_{z=0}=\bm v(\bm q)=v_\perp(\bm q)\hat{\bm{q}}_\perp$. The bulk Stokes equations (see Sec.~\ref{sec:SD}) give $P^\pm=\pm2\eta^\pm q \dot h(\bm q)\exp(\mp qz)$, $V_\parallel^\pm=-iqz\dot h(\bm q)\exp(\mp qz)$, $V_\perp^\pm=v_\perp(\bm q)\exp(\mp qz)$
and $V_z^\pm=(1\pm qz)\dot h(\bm q)\exp(\mp qz)$. One
can thus calculate the stresses transmitted by the bulk onto the membrane: $\bm\sigma^\pm=-\eta^\pm qv_\perp \hat{\bm q}_\perp$ and $\Sigma^\pm=-2\eta^\pm q\dot h$.
The dynamical equations of the membrane, i.e., eqs.~(\ref{eqtan})-(\ref{eqnor}), read then in Fourier space
\begin{eqnarray}
&&(2\eta q+\eta_2q^2)v_\perp(\bm q)=
\bm f' e^{-i\bm q\cdot\bm R}\cdot\hat{\bm q}_\perp,
\label{self}
\\
&&v_\parallel(\bm q)=0,
\label{self2}
\\
&&4\eta q\dot h(\bm q)=-(\kappa q^4+\sigma q^2)h(\bm q) 
+ Bq^2G(q) e^{-i\bm q\cdot\bm R},
\hspace{18pt}
\label{eqh}
\end{eqnarray}
where $2\eta=\eta^++\eta^-$, and where
\begin{equation}
\bm f' =\bm f-B\int \mathrm d^2r\,G(\bm r-\bm R)\bm\nabla \nabla^2h 
\label{effective_force}
\end{equation}
is the applied force reduced by the pull-back due to the membrane-inclusion coupling. The last equation, eq.~(\ref{eqh}), determines the deformation of the membrane produced by the dragged inclusion.

Let us consider a steady state with $\dot{\bm R}=v_p\bm e_x$, where $v_p$ is the constant  particle's velocity and $\bm e_x$ is the direction in which the force is applied, as in Refs.~\cite{Demery10PRL,Quemeneur14PNAS}. In the coordinate system comoving with the inclusion, we have $\bm R=\bm0$ and $\dot h(\bm r)$ becomes $-v_p\partial_xh(\bm r)$, so that eq.~(\ref{eqh}) gives
\begin{equation}
    h(\bm q)=\frac{BG(q)}{\kappa q^2+\sigma}\left(1+\frac{4i\eta q_xv_p}{\kappa q^3+\sigma q}\right)+\mathcal{O}(v_p^2).
\end{equation}
The effective force applied to the protein is then given by eq.~(\ref{effective_force}), yielding, at linear order in the velocity,
$f'=f-\gamma v_p$,
with
\begin{equation}
\gamma=2\eta B^2\!\int_{\bm q}
\frac{q^3 G(q)^2}{(\kappa q^2+\sigma)^2},
\label{voilagamma}
\end{equation}
where
$\int_{\bm q}=(2\pi)^{-2}\!\int_0^{\Lambda}\! {\rm d}^2q
$, with an upper wavevector cutoff $k_\mathrm{max}=\Lambda$ of the order of the inverse membrane thickness.
Injecting $\bm f'$ in eqs.~(\ref{self}) and (\ref{self2}) yields $\bm v(\bm q) = \mathbf O(\bm q)\cdot\bm f'$, where $\mathbf O$ is the Oseen-like tensor (\ref{modifiedoseen}) of the SD problem. Proceeding like in Sec.~\ref{sec:SD}, we thus obtain $v_p=\tilde\mu(f-\gamma v_p)$, where $\tilde\mu$ is the SD mobility (\ref{SDlaw2}). Hence, $f=(\tilde\mu^{-1}+\gamma)v_p$, so that the mobility $\tilde\mu_c$ for an inclusion curving the membrane is given by
\begin{equation}
\frac1{\tilde\mu_c}=\frac{1}{\tilde\mu}+\gamma.
\end{equation}
This implies, thanks to the Einstein relation, that the effective diffusion coefficient $D_\mathrm{eff}=k_\mathrm{B}T\tilde\mu_c$ is given in term of the bare diffusion coefficient $D_0=k_\mathrm{B}T\tilde\mu$ by
\begin{equation}
D_\mathrm{eff}
=D_0\left(1+\frac{D_0\gamma}{k_\mathrm{B}T}\right)^{-1}.
\end{equation}
This result coincides with that of Ref.~\cite{Quemeneur14PNAS}, with the correspondance $B=\frac12\kappa\Theta$ and $\Lambda=2\pi/a$ (see the eq.~(3) of the main text and the eq.~(S35) of the supporting information). The extra friction $\gamma$ comes from the mechanism introduced in Ref.~\cite{Demery10PRL}, which is the following. At rest, the inclusion sits in equilibrium at the top of the bump it creates. The pulling deforms the bump in such a way that the inclusion does not sit any longer at the minimum energy position: there is therefore a force that pulls it back; this force, the second term of eq.~(\ref{effective_force}), is responsible for the extra drag. The work produced by this drag is dissipated by the dynamics of the membrane deformation within the surrounding solvent. Indeed, as shown in Ref.~\cite{NajiPRL09}, this dynamics produces precisely the dissipation (\ref{voilagamma}).

As discussed in Ref.~\cite{Quemeneur14PNAS} using scaling arguments, the above correction to the diffusion coefficient yields $D_\mathrm{eff}\approx k_\mathrm{B}T/a$, when it is dominant, in agreement with Ref.~\cite{Demery10PRL} and with the Stokes-Einstein scaling law in $1/a$ obtained in Ref.~\cite{NajiBJ07}. Whether it may dominate and explain the experimental observations is disputable, however, as agreed by the authors of Ref.~\cite{Quemeneur14PNAS} and the authors of Ref.~\cite{Morris15PRL} who propose another mechanism based on an assumed tension-induced deformation of the protein shape.

\section{Discussion}
\label{Discussion}

We have shown that an excellent analytical approximation to SD law can be obtained very simply from the SD ``stokeslet" (the Oseen-like tensor of the SD problem) evaluated at the origin, upon regularizing it with an upper wavevector cutoff of the order of the inverse of the particle size $a$.
Using this method, we have investigated the consequences of the bilayer structure of the membrane (and of its asymmetry) and the role of the intermonolayer friction coefficient $b$.
We have also investigated the consequences of the deformation (bump) caused by a curvature-inducing particle.

In the case of an inclusion embedded in only one of the two monolayers, or simply adhering to one of them, we found that for large values of $b$ the SD law holds upon replacing the 2D viscosity $\eta_2$ of the membrane by the sum of the 2D viscosities of the monolayers. Indeed, $b$ can be neglected when it is large, as it effectively sets a no-slip boundary conditions between the two monolayers (they then act as an effective medium of viscosity $\eta_2=\eta_2^++\eta_2^-$). This breaks down when $b$ is smaller than $\eta_2/(4a^2)$, in which case the mobility gets larger since the monolayer opposite to the inclusion is not fully dragged by the inclusion around the latter.

In the  case of an inclusion spanning the whole bilayer, we found that for all practical values of $b$, the rule of replacing the 2D viscosity of the membrane by the sum of the monolayers viscosities holds. This is because the no-slip boundary condition between the inclusion and each of the two monolayers effectively imposes a no-slip  condition between the monolayers around the inclusion.

Finally, for curvature-inducing inclusions, we showed (in the small deformation regime) that the total friction is the sum of the SD friction and that due to the pull-back caused by the velocity induced  deformation of the bump. It would be interesting to investigate whether this remains true in a more general model involving a quadratic membrane-inclusion coupling.

\medskip
\textbf{Acknowledgments.} We thank P. Bassereau, H. Diamant, D. Lacoste, K. Mandadapu, N. Oppenheimer and F. van Wijland for useful discussions.


\begin{thebibliography}{42}

\bibitem{Lamb_book}
H.~Lamb, \emph{Hydrodynamics} (Cambridge University Press, New York, 1997)

\bibitem{SaffmanPNAS75}
P.G. Saffman, M.~Delbr{\"u}ck, Proceedings of the National Academy of Sciences
  \textbf{72}, 3111 (1975)

\bibitem{SaffmanJFM76}
P.G. Saffman, Journal of Fluid Mechanics \textbf{73}, 593 (1976)

\bibitem{HughesJFM81}
B.D. Hughes, B.A. Pailthorpe, L.R. White, Journal of Fluid Mechanics
  \textbf{110}, 349 (1981)

\bibitem{StoneJFM98}
H.A. Stone, A.~Ajdari, Journal of Fluid Mechanics \textbf{369}, 151 (1998)

\bibitem{StoneJFM15}
H.A. Stone, H.~Masoud, Journal of Fluid Mechanics \textbf{781}, 494 (2015)

\bibitem{Netz97JphysI}
R.R. Netz, J. Phys. I \textbf{7}, 833 (1997)

\bibitem{Park96JPhysI}
J.M. Park, T.C. Lubensky, J. Phys. I \textbf{7}, 1217 (1996)

\bibitem{Dommersnes99EPJB}
P.G. Dommersnes, J.B. Fournier, Eur. Phys. J. B \textbf{12}, 9 (1999)

\bibitem{Dommersnes02BiophysJ}
P.G. Dommersnes, J.B. Fournier, Biophys. J. \textbf{83}, 2898 (2002)

\bibitem{Bitbol10PRERC}
A.F. Bitbol, P.G. Dommersnes, J.B. Fournier, Phys. Rev. E \textbf{81},
  050903(R) (2010)

\bibitem{CortezSJSC01}
R.~Cortez, Siam J. Sci. Comput. \textbf{23}, 1204 (2001)

\bibitem{LevinePRE01}
A.J. Levine, T.C. Lubensky, Phys. Rev. E \textbf{63}, 041510 (2001)

\bibitem{PeskinAN02}
C.~Peskin, Acta Numerica \textbf{11}, 1 (2002)

\bibitem{AtzbergerJCP07}
P.J. Atzberger, P.R. Kramer, C.S. Peskin, Acta Numerica \textbf{224}, 1255
  (2007)

\bibitem{CamleyPRE11}
B.A. Camley, F.L.H. Brown, Phys. Rev. E \textbf{84}, 021904 (2011)

\bibitem{CamleySM13}
B.A. Camley, F.L.H. Brown, Soft Matter \textbf{9}, 4767 (2013)

\bibitem{SekiPRE14}
K.~Seki, S.~Mogre, S.~Komura, Phys. Rev. E \textbf{89}, 022713 (2014)

\bibitem{NajiPRL09}
A.~Naji, P.J. Atzberger, F.L.H. Brown, Phys. Rev. Lett. \textbf{102}, 138102
  (2009)

\bibitem{Quemeneur14PNAS}
F.~Quemeneur, J.K. Sigurdsson, M.~Renner, P.J. Atzberger, P.~Bassereau,
  D.~Lacoste, Proc. Natl. Acad. Sci. USA \textbf{111}, 5083 (2014)

\bibitem{Morris15PRL}
R.G. Morris, M.S. Turner, Phys. Rev. Lett. \textbf{115}, 198101 (2015)

\bibitem{MerkelJPhys89}
{Merkel, R.}, {Sackmann, E.}, {Evans, E.}, J. Phys. France \textbf{50}, 1535
  (1989)

\bibitem{Evans94CPL}
E.~Evans, A.~Yeung, Chem. Phys. Lipids \textbf{73}, 39 (1994)

\bibitem{Seifert93EPL}
U.~Seifert, S.A. Langer, Europhys. Lett. \textbf{23}, 71 (1993)

\bibitem{OtterBJ07}
W.~den Otter, S.~Shkulipa, Biophysical Journal \textbf{93}, 423 (2007)

\bibitem{FournierPRL09}
J.B. Fournier, N.~Khalifat, N.~Puff, M.I. Angelova, Phys. Rev. Lett.
  \textbf{102}, 018102 (2009)

\bibitem{SimunovicTCB15}
M.~Simunovic, G.A. Voth, A.~Callan-Jones, P.~Bassereau, Trends Cell Biol.
  \textbf{25}, 780 (2015)

\bibitem{NajiBJ07}
A.~Naji, A.J. Levine, P.~Pincus, Biophysical Journal \textbf{93}, L49  (2007)

\bibitem{CamleyPRE12}
B.A. Camley, F.L.H. Brown, Phys. Rev. E \textbf{85}, 061921 (2012)

\bibitem{Demery10PRL}
V.~Demery, D.~Dean, Phys. Rev. Lett. \textbf{104}, 080601 (2010)

\bibitem{DoiEdwards_book}
M.~Doi, S.F. Edwards, \emph{The Theory of Polymer Dynamics} (Clarendon press,
  Oxford, 1986)

\bibitem{LubenskyPF96}
D.K. Lubensky, R.E. Goldstein, Phys. Fluids \textbf{8}, 843 (1996)

\bibitem{Oppenheimer10PRE}
N.~Oppenheimer, H.~Diamant, Phys. Rev. E \textbf{82}, 041912 (2010)

\bibitem{Hill14PRCA}
R.J. Hill, C.Y. Wang, Proc. R. Soc. A \textbf{470}, 20130843 (2014)

\bibitem{Mathematica}
We used NIntegrate in Mathematica 12.0, Wolfram inc.

\bibitem{Gruler75ZN}
H.~Gruler, Z. NaturForsch. \textbf{C 30}, 608 (1975)

\bibitem{Leibler86JP}
S.~Leibler, J. Phys. \textbf{47}, 507 (1986)

\bibitem{Goulian93EPL}
M.~Goulian, R.~Bruinsma, P.~Pincus, EPL \textbf{22}, 145 (1993)

\bibitem{Prevost15NC}
C.~Pr\'evost, H.~Zhao, J.~Manzi, E.~Lemichez, P.~Lappalainen, A.~Callan-Jones,
  P.~Bassereau, Nature Comm. \textbf{6}, 8529 (2015)

\bibitem{VanderWel16SR}
C.~van~der Wel et~al., Scientific Reports \textbf{6}, 32825 (2016)

\bibitem{Helfrich73}
W.~Helfrich, Z. NaturForsch. \textbf{C 28}, 693 (1973)

\bibitem{Brochard75JPhys}
F.~Brochard, J.F. Lennon, J. Phys. France \textbf{36}, 1035 (1975)

\end{thebibliography}
\end{document}